\documentclass[article]{emulateapj}
\usepackage{epsfig}
\usepackage{natbib}
\usepackage{color}
\usepackage{graphicx}

\def \etal {et~al.~}
\def \chisq  {\ifmmode  \chi^2   \else  $\chi^2$  \fi}  
\def \spose#1{\hbox  to 0pt{#1\hss}}  
\def \lta{\mathrel{\spose{\lower 3pt\hbox{$\sim$}}\raise  2.0pt\hbox{$<$}}}
\def \gta{\mathrel{\spose{\lower  3pt\hbox{$\sim$}}\raise 2.0pt\hbox{$>$}}}
\def \kms {\ifmmode  \,\rm km\,s^{-1} \else $\,\rm km\,s^{-1}  $ \fi }
\def \kpc {\ifmmode  {\rm kpc}  \else ${\rm  kpc}$ \fi  }  
\def \Msun {\ifmmode M_{\odot} \else $M_{\odot}$ \fi} 
\def \hMsun {\ifmmode h^{-1}\,\rm M_{\odot} \else $h^{-1}\,\rm M_{\odot}$ \fi}

\shorttitle{Interation between DM sub-halos $\&$ galactic disk}
\shortauthors{Kannan, Macci\`{o}, Pasquali, Moster $\&$ Walter}

\begin{document}

\title{Interaction between dark matter sub-halos $\&$ galactic gaseous disk}

\author{Rahul Kannan\altaffilmark{1}$^*$} \author{Andrea V. Macci\`{o}\altaffilmark{1}} \author{Anna Pasquali\altaffilmark{2}} \author{Benjamin P. Moster\altaffilmark{3}} \author{Fabian Walter\altaffilmark{1}}

\altaffiltext{1}{Max-Planck-Institut f\"{u}r Astronomie, K\"{o}nigstuhl 17, 69117 Heidelberg, Germany}
\altaffiltext{2}{Astronomisches Rechen Institut, Zentrum f\"ur Astronomie der Universit\"at Heidelberg, M\"onchhofstrasse 12 - 14, 69120 Heidelberg, Germany}
\altaffiltext{3}{Max-Planck-Institute for Astrophysics, Karl-Schwarzschild-Str. 1, 85741 Garching, Germany}

\email{*email:kannan@mpia.de}
 
\begin{abstract}

We investigate the idea that the interaction of Dark Matter (DM) sub-halos with the gaseous disk of galaxies can be the origin for the observed holes and shells found in their neutral hydrogen (HI) distributions. 
We use high resolution hydrodynamic simulations to show that pure DM sub-halos impacting a galactic disk are not able to produce holes; on the contrary, they result in high density regions in the disk. 
However, sub-halos containing a small amount of gas (a few percent of the total DM mass of the sub-halo) are able to displace the gas in the disk and form holes and shells. 
The size and lifetime of these holes depend on the sub-halo gas mass, density and impact velocity. A DM sub-halo of mass  $10^8$ $M_{\odot}$ and a gas mass fraction of $\sim 3$$\%$, is able
to create a kpc scale hole, with a lifetime similar to those observed in nearby galaxies. We also register an increase in the star formation rate at the rim of the hole, again 
in agreement with observations. Even though the properties off these simulated structures resemble those found in observations we find that the number of predicted holes (based on mass and orbital distributions 
of DM halos derived from cosmological N-body simulations) falls short compared to the observations. Only a handful of holes are produced per Giga year. This leads us to conclude that DM halo impact is not the major channel through which these holes are formed.
 
\end{abstract}


\section{Introduction}

Holes and shells have been found in the neutral hydrogen (HI) distribution of many nearby galaxies like M31 (\citealt{1981A&A....95L...1B}, \citealt{1986A&A...169...14B}), M33 
\citep{1990A&A...229..362D}, Holmberg II \citep{1992AJ....103.1841P}, M101 and NGC 6946 \citep{1993nhns.book.....K}, IC 10 \citep{1998AJ....116.2363W},    
SMC \citep{1997MNRAS.289..225S} , LMC \citep{1998ApJ...503..674K} and IC 2574 \citep{1999AJ....118..273W}. These structures are typically regions of low
HI density distributed across the entire galaxy, and come in various sizes 
ranging from a few 100 pc to about 1.5 kpc. Their expansion velocity sets
an upper limit to their dynamical age of about 10 - 60 Myr (cf.  \cite{1999AJ....118..273W} for IC 2574).

The formation of HI holes and shells has often been ascribed to the action
of stellar winds and supernova explosions occurring in OB associations
and young stellar clusters (cf. \cite{1977ApJ...218..377W}, \cite{1988ARA&A..26..145T}, \cite{1996ASPC..106...47V}). Based on a simple model of holes
created by O and B stars, \cite{1997MNRAS.289..570O} successfully predicted
the observed number distribution of holes in the SMC, lending support
to this hypothesis. More recently, the analysis conducted by \cite{2001AJ....122.3070O}, \cite{2005AJ....129..160S}, \cite{2009ApJ...704.1538W} and \cite{2011ApJ...735...36C} have shown that the stellar content of HI holes in Holmberg I and II, DDO 88 and 165 can release enough mechanical energy to drive the formation of holes on timescales similar to what is observed. In particular, \cite{2009ApJ...704.1538W} were able to show that all the holes in Holmberg II observed with HST
contain multiple stellar populations of different age.  \cite{2011arXiv1105.4117W} studied  the HI holes in five dwarf galaxies (DDO 181, Holmberg I, 
M81 Dwarf A, Sextans A and UGC 8508), and failed to detect a young star 
cluster at the center of the observed HI holes. They thus suggested
that large holes may form due to multiple episodes of star formation. 

There is more observational evidence against a single burst of star formation
being responsible for the creation of HI holes. For example, no robust
spatial correlation was found by \cite{1999AJ....118.2797K} between the distribution of the H$\alpha$ emission and the HI holes in the LMC. \cite{2005MNRAS.360.1171H} estimated that the holes in the SMC not associated with a young star 
cluster are a factor of 1.5 more than those exhibiting relatively young stars 
at their center. In the Galaxy, the distribution of radio pulsars compared 
to that of holes seem to indicate that holes can not be produced by the 
supernova explosions of a single age stellar population \citep{2004ApJ...606..326P}. \cite{2008ApJ...687.1004P} found that the most recent episode of star
formation has taken place preferentially at the rims of the HI holes in
IC 2574, exception made for the supergiant shell which indeed embeds a young
star cluster (cf. \cite{2000AJ....120.1794S}, \cite{2009ApJ...691L..59W}). 
\cite{1999AJ....118..323R} observed the HI holes in Holmberg II in search of the
descendants (stars of spectral types B, A and F) of those clusters whose
supernovae would have produced the holes. They measured an integrated light
of the stars within the holes inconsistent with the hypothesis of a young
star cluster triggering the formation of a HI hole. In the case of IC 1613,
its largest HI shell was found to host about 27 OB associations whose
energy can not sustain the formation and expansion of the shell (\cite{2004A&A...413..889B}, \cite{2006A&A...448..123S}).

Some authors have proposed alternative formation hypotheses to the SNe origin. 
\cite{1998ApJ...501L.163E} and \cite{1998ApJ...503L..35L} postulate that a high-energy gamma ray burst (GRB) from the death of a single massive star could create 
kpc sized holes in the Inter Stellar Medium (ISM), thus offering an explanation for holes without a detectable underlying cluster. These authors assume the energy from GRBs is emitted isotropically. 
However, GRBs release most of their energy in bi-directional beams (eg., \citealt{1977MNRAS.179..433B} mechanism), making this scenario less likely to produce large HI holes.

Another mechanism proposed is the infall of gas clouds \citep{1987A&A...179..219T}. One observational prediction of this model is a half-circle arc seen in an 
HI position-velocity diagram. The half-circle arc arises from the gas being pushed to one direction, 
corresponding to the direction of travel of the high velocity clouds. Some observational support for this idea is reported by Heiles (\citealt{1979PASP...91Q.611H}, \citealt{1984ApJS...55..585H}) 
who point out that the most energetic Galactic shells in their study all have half-circle arc signatures in position-velocity space ( \citealt{1991A&A...244L..29K}).

In this paper we investigate an alternative formation scenario, where holes and shells in extended HI disks are the result of interaction of the
gaseous disk with dark matter (DM) sub-halos.
The Lambda Cold Dark Matter (LCDM) model predicts the existence of thousands of dark matter substructures within the dark matter halo of every galaxy
(e.g. \citealt{2008MNRAS.391.1685S} and reference therein). The majority of these sub-halos is ``dark'' i.e. does not host a satellite galaxy or any visible stellar
structures (e.g. \citealt{2010MNRAS.402.1995M}, \citealt{2011arXiv1102.2526F}).
If these sub-halo population is able to produce the observed holes in galactic HI disks, like the one of IC 2574, then this might provide a new way of 
detecting the presence of non-luminous DM halos. For example, the number of holes could then be linked statistically to the amount of substructure 
in the dark matter distribution  and thus provide a unique way to asses the nature of dark matter and by extension to test the $\Lambda$CDM model.

Perturbations on a stellar disk due to (massive) satellites have been extensively studied in the recent years (e.g. \citealt{2008ApJ...688..254K}, \citealt{2010MNRAS.403.1009M} and references
therein), while less attention has been given to the effect of low mass DM clumps on an extended gaseous disk, with few exceptions.
\cite{2006ApJ...637L..97B}  investigate how the impact of dark matter sub-halos orbiting a gas-rich disk galaxy embedded in a massive dark matter halo influences the 
dynamical evolution of the outer HI gas disk of the galaxy. They show that the impact of dark matter sub-halos (``dark impact'') can be important for better understanding 
the origin of star formation discovered in the extreme outer regions of disk galaxies.  They also discuss the possibility that this kind of dark impact will be able to produce holes in 
the gas distribution. In their study they adopted a model, which did not include multiple events, a live dark matter halo (they adopted an analytic fixed potential),
cooling and star formation.

A new attempt to detect the imprint of the dark satellites in the HI disk of the Milky Way has been recently made. In a series of papers \cite{2009MNRAS.399L.118C} and \cite{2011ApJ...731...40C} 
examine tidal interactions between perturbing dark sub-halos and the gas disk of the Milky Way using high-resolution Smoothed Particle Hydrodynamics simulations.
They compare their numerical results to the observed HI map of the Milky Way, and find that the Fourier amplitudes of the planar disturbances are best fitted by a perturbing 
dark sub-halo with a mass that is one-hundredth of the Milky Way with a pericentric distance of 5 kpc. 
More recently  ( \citealt{2011arXiv1102.3436C}) develop a perturbative approach to study the excitation of disturbance in the extended atomic hydrogen discs 
of galaxies produced by  passing dark matter sub-halos. They show that the properties of dark matter sub-halos can be inferred from the profile and amplitude of the 
different perturbed energy modes of the disk.

Motivated by these recent studies, in this paper we use high resolution hydro-dynamical simulations to study in details the interaction of DM sub-halos with a galactic gaseous disk.
Our primary goal is to see under which conditions DM satellites are able to create holes that resemble the observed ones and whether the majority of these holes can be 
explained by dark satellites-disk interactions. We model our primary galaxy on the nearby dwarf galaxy IC 2574 and try to replicate its observed features in our simulations. 
After having described our numerical setup, we will first present results of a single satellite-disk interaction, for different satellite orbital parameters and gas content. 
Then we will use satellite parameters (mass, velocity and position) directly obtained from high resolution N-body simulations to study the frequency of DM sub-halo disk
encounters. Finally we will discuss the implications of our results.

\section{Numerical Simulations}
\label{sec:sim}

We make use of the parallel TreeSPH-code GADGET-2 \citep{2005MNRAS.364.1105S} in this work. 
The code uses SPH (\citealt{1977MNRAS.181..375G}; \citealt{1977AJ.....82.1013L}; \citealt{1992ARA&A..30..543M}) to evolve the gas using an entropy conserving scheme \citep{2002MNRAS.333..649S}. Radiative
cooling is implemented for a primordial mixture of hydrogen and
helium following \cite{1996ApJS..105...19K} and a spatially uniform time-independent local ultraviolet background in the
optically thin limit \citep{1996ApJ...461...20H} is included.
The SPH properties of the gas particles are averaged over the
standard GADGET-2 kernel using 64 SPH particles. Additionally,
the minimum SPH smoothing length is required to be equal to the
gravitational softening length in order to prevent artificial stabilization of small gas clumps at low resolution \citep{1997MNRAS.288.1060B}.

All simulations have been performed with a high force accuracy of
$\alpha_{force} = 0.005 $ and a time integration accuracy of $\eta_{acc} = 0.02 $ (for
further details see \citealt{2005MNRAS.364.1105S}).
Star Formation (SF) and the associated heating by supernovae (SN) are modeled
following the sub-resolution multiphase interstellar medium (ISM)
model described in \cite{2003MNRAS.339..289S}. The ISM in the
model is treated as a two-phase medium with cold clouds embedded
in a hot component at pressure equilibrium. Cold clouds form stars
in dense $(\rho > \rho_{th} )$ regions on a time-scale chosen to match observations \citep{1998ApJ...498..541K}. The threshold density $\rho_{th}$ is determined
self-consistently by demanding that the equation of state (EOS) is
continuous at the onset of SF. We do not include feedback from accreting black holes (AGN feedback) in
our simulations as there is no evidence of AGN activity in dwarf galaxies like the one studied here.
In our runs, the parameters for the star formation and feedback are adjusted to match the Kroupa initial mass function (IMF) as specified by \cite{2001MNRAS.322..231K}. 
The SF time-scale is set to $t^0_* = 3.5$ Gyr, the cloud evaporation parameter to $A_0 = 1250$
and the SN temperature to $T_{SN} = 1.25\times 10^8 K$.\\

\subsection{Primary Galaxy Setup}
We apply the method given by \cite{2005MNRAS.361..776S} to construct the central galaxy. The central galaxy consists of an exponential stellar disk. In order to match the almost constant radial gas density profile of IC 2574, we have modeled the gas disk by two components. A radial exponential component and a constant radial profile which falls off rapidly at a specified scale radius. The stellar disk has a mass $M_{\rm {disk}}$, the gaseous disk has a mass $M_{\rm{gas}}$, with a spherical bulge with mass $M_{\rm{b}}$ embedded in a dark matter halo of mass $M_{\rm{vir}}$. The halo has a \cite{1990ApJ...356..359H} profile with a scale radius corresponding to a Navarro
Frenk $\&$ White halo (NFW; \citealt{1997ApJ...490..493N}) with a
scale-length of $r_{\rm{s}}$ and a concentration parameter $c_{\rm{vir}} = R_{\rm{vir}} /r_{\rm{s}}$. We use
the results of \cite{2008MNRAS.391.1940M} to compute
halo concentration as a function of virial mass.
The scale-lengths $r_d$ of the exponential gaseous and stellar disks
are assumed to be equal, and are determined using the model of \cite{2008MNRAS.391.1940M}, assuming that the fractional angular
momentum of the total disk $j_{\rm{d}} = (J_{\rm{gas}} + J_{\rm{disk}})/J_{\rm{vir}}$ is equal to the
global disk mass fraction $m_{\rm{d}} = (M_{\rm{gas}} + M_{\rm{disk}})/M_{\rm{vir}}$ for a constant
halo spin $\lambda$. This is equivalent to assuming that the specific angular
momentum of the material that forms the disk is the same as that of
the initial dark matter halo, and is conserved during the process of
disk formation.
The vertical structure of the stellar disk is described by a radially
independent $sech^2$ profile with a scale-height $z_0$ , and the vertical velocity dispersion is set equal to the radial velocity dispersion. The
vertical structure of the gaseous disk is computed self-consistently
as a function of the surface density by requiring a balance of the
galactic potential and the pressure given by the EOS. The stellar
bulge is constructed using the \cite{1990ApJ...356..359H} profile with a scale-
length $r_b$ .

\begin{figure}
\begin{center}
\epsscale{1.20}
\epsfig{file=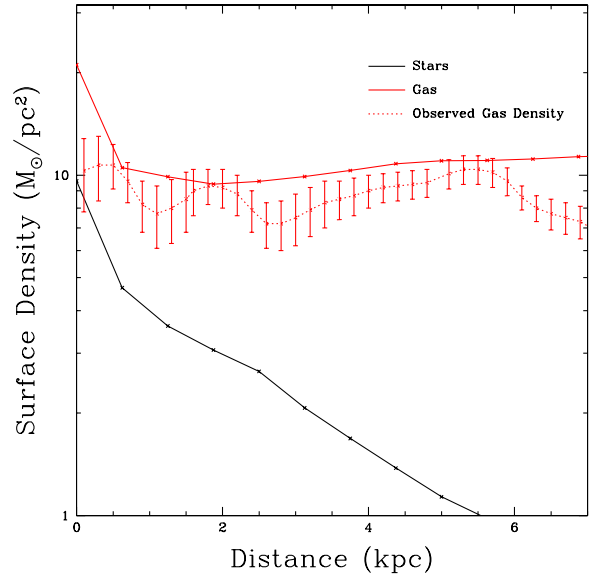, scale=0.4}
\caption{The red line is the gas surface density and the black line is the surface density of stars in the inner $7$ kpc of IC 2574. The observational data is taken from Figure 27 of \citep{2008AJ....136.2782L}. The error bars represent the $\rm{rms}$ uncertainty value.}
\label{srdens}
\end{center}
\end{figure}

We build a primary galaxy which matches the properties of IC 2574 (mass profiles taken from \citealt{2008AJ....136.2782L}). The galaxy has a halo of 
mass $M_{\rm{200}}= 10^{11} M_{\odot}$ obtained from the stellar mass, from the recipe given in \cite{2010ApJ...710..903M}, containing a stellar 
disk of mass $M_{\rm{disk}} = 3.16 \times 10^8 M_{\odot}$, a gaseous disk component of mass  $M_{\rm{gas}} = 1.8 \times 10^9 M_{\odot}$
 with $80$$\%$ of the gas in the HI disk, a small bulge of mass $M_{\rm{bulge}} = 3.16 \times 10^7 M_{\odot}$. The DM halo has a 
concentration parameter of $c_{\rm{200}} = 8.47$. The scale radius of the disk (stellar $\&$ the exponential component of the 
gaseous disk) is $r_{\rm{d}}=2.1$ kpc and the scale radius of the slab of gas is $r_{\rm{slab}}=4.5$ kpc. Once we fix $r_{\rm{d}}$ 
we compute the halo spin parameter from the recipe of  \cite{1998MNRAS.295..319M} which results in a value of $\lambda = 0.047$. 
The disk scale-height is fixed at $z_0 = 0.15$ kpc and scale radius of the bulge is set at $r_{\rm{b}} = 0.095$ kpc. The galaxy 
has $N_{\rm{DM}} = 10^6$ DM particles, $N_{\rm{disc}} = 10^5$ in the stellar disc, $N_{\rm{bulge}}= 10^4$ in the bulge, and $N_{\rm{gas}} = 5.7 \times 10^{5}$ 
gaseous disk particles. 
The force resolution (softening) is 101,80,57 pc for dark, gas, and stars respectively
In order to have a stable initial condition we evolve this primary galaxy in isolation for $1$ Gyr. 
The initial conditions were chosen in such a way that the surface density of the gas match the observed density of IC 2574 after evolving 
for $1$ Gyr as shown in Figure \ref{srdens}.

\section{Simulation Results}
\label{sec:res}

We here present the results of our numerical experiments. We mainly run three kinds of simulations: single encounter with a pure DM sub-halo (no gas or stars);
single encounter with a DM sub-halo containing a small gas fraction; and multiple, cosmologically motivated encounters. 

These single halo simulations are simplistic experiments designed to explore the effect of sub-halo/disk interactions. 
Orbit, mass and gas content are not meant to represent the typical case. An cosmological motivated run with multiple interactions will be presented in section \ref{ssec:cosmo}.

\subsection{Pure DM Sub-halo Interaction}
\begin{figure}
\begin{center}
\epsfig{file=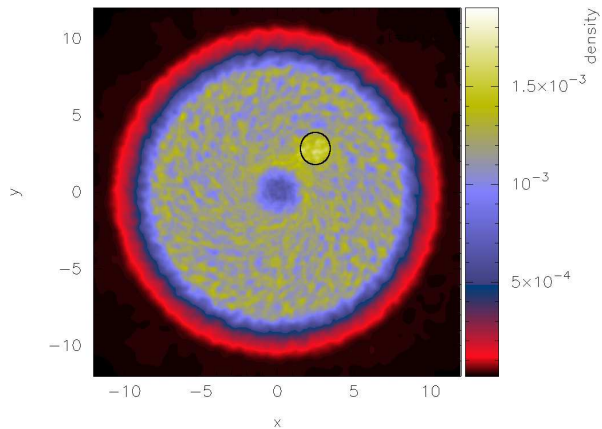, angle=270, scale=0.3}
\caption{Surface density map of the gaseous disk after the passage of the pure DM sub-halo of mass $10^8 M_{\odot}$. Such an encounter leads to a gas overdensity (marked by a circle). The units of density in this plot are in $10^{10} M_{\odot}kpc^{-2}$. The 2D contour plots were made using the visualization software SPLASH (\citealt{2007PASA...24..159P}).}
\label{sdpdm}
\end{center}
\end{figure}

In this section we investigate the dynamical impact of a pure DM halo on the gaseous disk. The mass of the sub-halo is first fixed at a value of 
$M_{\rm{sb}} = 10^8 M_{\odot} $.  
We want to have a highly concentrated sub-halo to act like a 'bullet' in the interaction. To do this we start with a $M_{200} = 10^{10} M_{\odot}$ halo and then carve out $99\%$ of its mass and create a $10^8 \Msun$ halo. This is justified, as it has been shown that the sub-halos passing through the primary halo of a galaxy get tidally stripped and can lose up to $90\%$ and in some extreme cases $99\%$ of their mass during their orbit (see \citealt{2008ApJ...673..226P}; 
\citealt{2010MNRAS.402.1995M}). The total number of DM particles in halo has been set to $N_{DM} = 1000$.
Starting from a lower mass halo (e.g $10^9 \Msun$) would have resulted in a less concentrated 'bullet', hence our choice maximizes the dynamical effect on the disk.
This sub-halo is  placed at a distance of $5$ kpc above the gaseous disk of the primary galaxy and given an initial velocity of $v_z=150$ kms$^{-1}$ perpendicular and pointing towards the galactic disk. 
Figure \ref{sdpdm} shows a surface density map of the gaseous disk after the passage of the sub-halo.

As it emerges clearly from Figure \ref{sdpdm}, a pure DM sub-halo is unable to form a hole, instead it gives rise to a localized high density region (marked by a circle). 
This is due to the fact that the DM particles are collision less. 
Due to the lack of contact forces the sub-halo cannot push away the gas particles. 
DM can only gravitationally focus the gas  in a stream behind its path of motion (Bondi-Hoyle accretion;\citealt{2004NewAR..48..843E};\citealt{1944MNRAS.104..273B};
\citealt{1952MNRAS.112..195B} ) which gives rise to higher surface density in the region it passes through. 
These high density peaks are $20\%-25\%$ denser than the mean density of the disk, and they last for about $80-90$ Myr, 
after wards they are destroyed by the dispersion of the gas and the differential rotation of the disc.
The fundamental result of this first experiment is that a pure dark matter sub-halo is not able to create a hole in the gaseous disk.

\subsection{Gaseous Sub-Halo Interaction}

Since a pure DM halo is not able to displace the gas in the disk, a medium which interacts through gravity as well as contact forces is needed in order to push the gas away 
from its motion path. A gaseous medium will provide the required contact forces to push the gas away and form holes. 
We construct the DM sub-halo with the same mass and concentration parameter as described in the previous section and 
in addition we also add a hot gaseous component. The hot gas has a beta profile with $\beta = 2/3$, a spin factor of 
$\alpha=4$ and a core radius of $90$ pc. For a more detailed description of the hot gas profile and parameters we refer 
to \cite{2011arXiv1104.0246M}. These sub-halos are expected to have a very low amount of gas (if any), due to re-ionization (e.g. \citealt{2008MNRAS.390..920O}).
\footnote{Actually since we started from a $10^{10} \Msun$ halo, our halo could possibly contain some stars, but in this experiment we 
decided to neglect this component, since it won't change the overall picture.}

Moreover we are interested in ``dark'' satellites, so we do not want the gas in the sub-halos to form stars, 
hence we choose a very low gas fraction (i.e. fraction of mass in gas relative to the DM mass) in these sub-halos: $M_{gas}/M_{DM}\approx 0.03$. 
We consider only an hot gas component; as halos containing cold gas (and possibly stars), will be directly detectable; while, in this work, we are 
interested in testing the effects of the more numerous, undetected, dark satellite population, which is predicted by the cold dark matter model.
We then place this  new sub-halo on the exact same orbit of the pure dark matter experiment and let it interact with the gaseous disk as described in the previous section.

\begin{figure}
\begin{center}
\epsfig{file=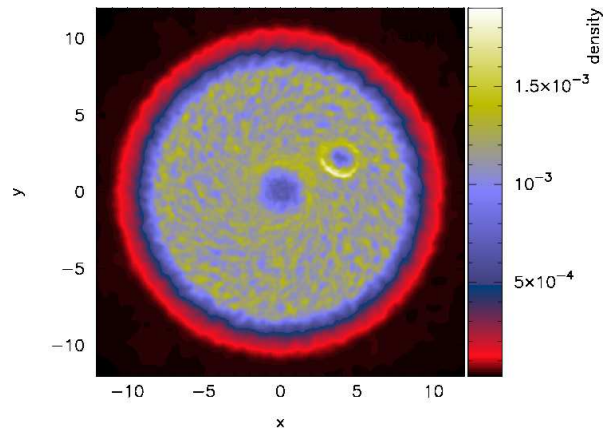, angle=270, scale=0.3}
\caption{Effect of the interaction between a DM sub-halo containing $3.2\%$ 
of gas by mass(DM mass = $10^8 M_{\odot}$). }
\label{sd3pnt2wns}
\end{center}
\end{figure}

The hot gas in the halo is able to displace the gas in the disk and produce a low density region (a hole), associated with an expanding high density shock wave as shown in Figure \ref{sd3pnt2wns}.
For our chosen parameters the hole has a diameter of about $1.5$ kpc and lasts for about $60 $ Myr.
The lifetime of the holes is decided by two factors, the pressure gradient  and the differential rotation. 
Our simulations suggest that the pressure gradient causes the hole to be filled by gas long before it gets destroyed by the differential rotation, 
although differential rotation becomes important for very large holes. For typical hole sizes of $1-2$ kpc  we find an average hole lifetime of about 50-60 Myr; in agreement with the dynamical ages estimated by \cite{1999AJ....118..273W}.

\begin{figure}
\begin{center}
\epsfig{file=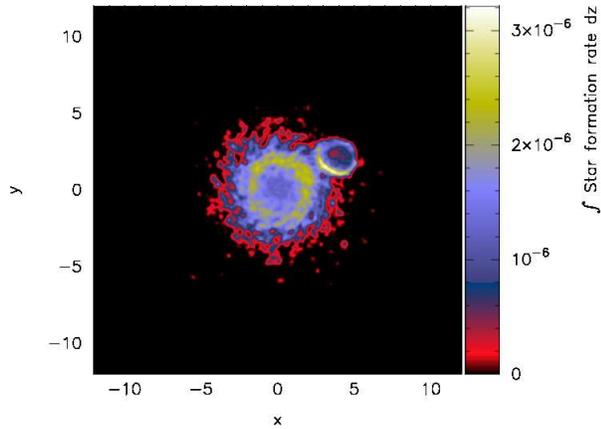, angle=270, scale=0.3}
\caption{Star formation rate, showing clearly the ring of new stars formed along the rim of the hole. The units of SFR surface density in this plot are in $10^{10} M_{\odot}kpc^{-2}$.}
\label{sfr}
\end{center}
\end{figure}

Another interesting feature of this simulation is the enhanced star formation
on the rim of the hole, as shown in Figure \ref{sfr}. New stars have preferentially
formed on the high density edges of the hole, while the hole itself contains
very few new stars. This result is in agreement with the ages of the stars
associated with the HI holes in IC2574 as derived by \cite{2008ApJ...687.1004P},
via comparison of observed UBV colors with those predicted by synthetic
stellar populations (for more details see \citealt{2008ApJ...687.1004P}).

\begin{figure}
\begin{center}
\epsfig{file=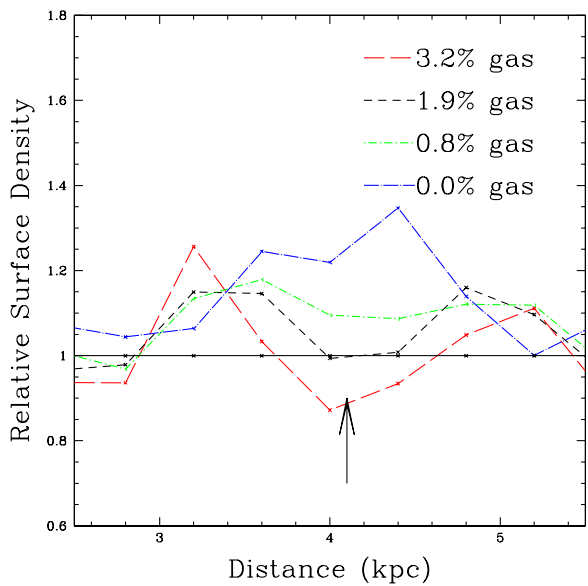, scale=0.4}
\caption{The effect of the interaction between DM sub-halos with different gas fractions and the disk  
The curves denotes the relative surface density variation with respect to the isolated disk, upon the interaction with a halo with $3.2$$\%$ gas (red), $1.9\%$ gas (green), $0.8\%$ gas (blue) and $0\%$ gas (black). The arrow indicates to point of impact of the halos. (DM mass = $10^8 M_{\odot}$)}
\label{sd3pnt2dv1}
\end{center}
\end{figure}

We then run a series of simulations for different percentages of gas fraction of the DM sub-halos,  
keeping all other simulation parameters constant. The effect of a pure DM sub-halo and a sub-halo containing gas are inherently different. 
Figure  \ref{sd3pnt2dv1} shows the relative surface density profiles of the gas along a strip of width $1$ kpc joining the center of the galaxy and the center of the hole, for different
sub-halo gas fractions (including the case of no gas).
A pure DM sub-halo (in black) produces a peak of increased density while a gaseous halo produces a low density region surrounded by high density wave. 
This density wave increases the density in the rim of the hole above the star formation density threshold ($\rho_{\rm{th}}$), thus triggering star formation all along the rim of the hole. As expected, the size of the hole and the extent to which the disk is perturbed depends on the gas fraction of the infalling DM sub-halo.

\begin{figure}
\begin{center}
\epsfig{file=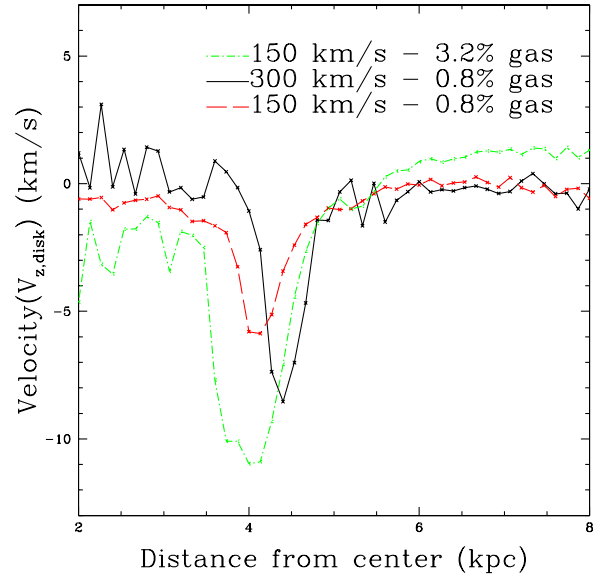, scale=0.4}
\caption{Variation of $V_{z,disk}$ with initial velocity. The red curve traces the change in the velocity profile in the $z$ direction for a halo which has $0.8\%$ gas  and velocity of $v_z = 150$ km s$^{-1}$ and the black line for $v_z=300$ km s$^{-1}$, the green curve is for a gas fraction of $3.2\%$ and velocity of 150 km s$^{-1}$.}
\label{sd3pnt2dv}
\end{center}
\end{figure}

The impact with a sub-halo will create a net velocity gradient along the sub-halo trajectory in the vertical component of the H I gas. This signature could in principle disentangle 
different formation scenarios for the HI holes. Figure  \ref{sd3pnt2dv} shows the velocity in the $z$-component (perpendicular to the disk plane - $V_{z,disk}$ which equals zero in the unperturbed case) along 
the same strip as mentioned above. Here the infalling halos has two different gas fractions, $3.2\%$ (green line)  $\&$ $0.8\%$ (red line) with a velocity of $v_z=150$ km s$^{-1}$ and a halo with a gas fraction of $0.8\%$ with a veloocity of $v_z=300$ km s$^{-1}$. A higher sub-halo initial velocity and/or a larger gas fraction imparts more energy to the disk in the impact direction and hence produces a stronger signature in $v_{z,disk}$. This plot shows that the effect is small and would be very hard to detect in real observations as the effect is of the same order as the typical velocity resolution of HI observations of nearby galaxies (e.g. \citealt{2008AJ....136.2563W})

\subsection{Cosmological runs}   
\label{ssec:cosmo}

Our previous simplified setups have shown that dark matter sub-halos that contain a small fraction of gas can in principle create holes in the HI distribution that resemble the observed ones.
The next question to answer now is whether there are enough DM satellites to reproduce the number of observed holes in a galaxy similar to IC2574.

To answer this question we turn to cosmological N-body simulations which give us the mass, size, velocity and orbital parameters of sub-halos of the reference galaxy IC2574.
We use the publicly available database of satellite distribution in the Via Lactea II (VLII) simulation (\citealt{2007ApJ...667..859D}, 2008).
Unfortunately the mass of the parent halo in the VLII simulation ($2\times10^{12} M_{\odot}$) is higher than the estimated mass of the halo of our reference galaxy IC2574 ($10^{11} M_{\odot}$).
We thus decide to scale down the properties of the sub-halos using the following simple dynamically motivated scaling relations:
\begin{equation}
M^{\prime} = M/10 
\end{equation}
\begin{equation}
v^{\prime}_{x,y,z} = \frac{v_{x,y,z}}{10^{1/3}}
\end{equation}
\begin{equation}
x^{\prime} = \frac{x}{10^{1/3}} \;\;\;\; y^{\prime} = \frac{y}{10^{1/3}} \;\;\;\; z^{\prime} = \frac{z}{10^{1/3}}.
\end{equation}

\begin{figure}
\begin{center}
\epsscale{1.20}
\epsfig{file=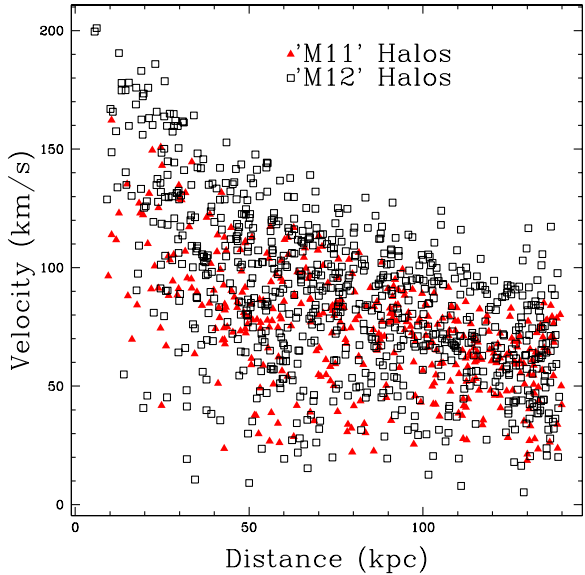, scale=0.4}
\caption{The velocity as a function of distance for the M12 sub-halos(black open squares) after being scaled down and M11 ones (red triangles). }
\label{dispvel}
\end{center}
\end{figure}

In order to test that we have not introduced any biases in the satellite properties by rescaling them, we run an additional N-body simulation of a $M=10^{11} M_{\odot}$ dark matter halo, at a 
resolution lower than the VLII. 
We have selected  the candidate halo from
an existing  low resolution dark matter  simulation (350$^3$ particles
within 90  Mpc, see \cite{2010MNRAS.403..984N}) 
and re-simulated  them at higher resolution  using the
volume  renormalization  technique \citep{1993ApJ...412..455K}.
The total number of particles within the virial radius at $z=0$ is $\approx 3.3 \times 10^6$, 
which gives a mass per particle of $3.05 \times 10^4 \Msun$.
We will refer to this simulation as M11, while we will use M12 for the rescaled version of the VLII simulation.
Figure \ref{dispvel} shows that sub-halos of both runs  occupy the same phase-space region, which confirms that our simple scaling is valid as first order
approximation.\footnote{The shelf of high velocity subhalos at a given distance
is due to the higher halo statistics in M12, that allows a better sampling
of the tails (positive and negative) of the velocity distribution.} We use both satellite distributions (form M11 and M12) in our cosmologically motivated tests.

These N-body simulations predict that a large number of sub-halos are present around a typical galaxy of $10^{11} M_{\odot}$. 
In fact we have 798 sub-halos from the M12 simulation and 468 from the M11 simulation, with masses larger than $10^6 M_{\odot}$, our resolution limit in the N-body simulations. 
Not all of these halos pass through the disk. Most of them have nearly circular orbits. Only a few of them have orbits which take them close enough to the center of the halo, and hence interact with the disk,
these are the halos we are interested in.
We find interacting halos by integrating orbits of all sub-halos in a static potential for 1 Gyr.
This static potential is parametrized with an NFW profile, with the same virial radius and concentration of the original halo extracted from the Nbody simulation at $z=0$.
\footnote{We use as starting point the Nbody results at $z_i=0$. Our integration time of 1 Gyr would in principle require
$z_i=0.079$. Such a tiny difference should not affect the sub-halo mass function on the small scales as considered in this work ($M\approx 10^7$).}.
A interaction is defined as the passage of the sub-halo within a cylinder of radius $R=\sqrt{x^2+y^2}<8$ kpc and a height $-2<h_z<2$ centered at the center of the halo. 
We obtain a sample of 37 dark matter sub-halos for M12 and 13 for M11, where the lower number of satellites in the M11 run is due to the its lower resolution compared to 
the VLII simulation.

\begin{center}
\begin{deluxetable}{cccccc}
\tabletypesize{\large}
\tablecaption{Number of halos and their properties}
\tablewidth{0pt}
\tablehead{
\colhead{Mass($M_{\odot}$)}
& \colhead{$N_{H}-M12$}
& \colhead{$N_{H}-M11$}
& \colhead{$c_{200}$}
& \colhead{$N_{DM}$}
& \colhead{$N_{Gas}$}
}
\startdata
$5.56 \times 10^6$ & 26 & 10 & 44.16 & 556  & 446  \\
$1.01 \times 10^7$ & 7  & 0  & 41.65 & 1010 & 799  \\
$1.46 \times 10^7$ & 3  & 2  & 40.16 & 1463 & 1158 \\
$1.91 \times 10^7$ & 0  & 0  &   -   &  -   & -    \\
$2.37 \times 10^7$ & 1  & 1  & 38.34 & 2371 & 1876 \\
\enddata 
\tablecomments{The first column denotes $M_{200}$ in units of solar mass, the second, number of halos in M12 run, the third, number of halos in M11 run, the fourth, concentration parameter, the fifth, number of DM particles used to sample the halo and sixth, number of gas particles used to sample the halo}
\label{table1}
\end{deluxetable}
\end{center}

In both the M12 and M11 runs DM sub-halo masses range between $1.1 \times 10^6 M_{\odot} - 1.3 \times 10^7 M_{\odot}$. 
We divide these halos into five halo mass bins. The number of particles in each mass bin and their properties are given in Table \ref{table1}. 
All halos in this run have a gas fraction $5\%$ by mass. We put this relatively large amount of gas in these halos so as to get an upper limit to the number of holes which can be formed.

For each of the mass bins we generate a synthetic halo, using a recipe, which is slightly different from the single halo simulation presented in the previous section.
We create a spherical NFW halo with a density contrast of $\Delta=1000$ with respect to the critical density of the universe, as is typical for sub-halos 
(the density contrast for isolated virialized halos is normally assumed to be $\Delta_{vir} \approx 100$).
Each of these halos contains a gaseous halo in hydrostatic equilibrium with the DM potential.
Finally we place back our synthetic halos around the primary galaxy at the same position and with the same velocity as obtained from the N-body simulations.

\begin{figure}
\begin{center}
\epsscale{1.00}
\epsfig{file=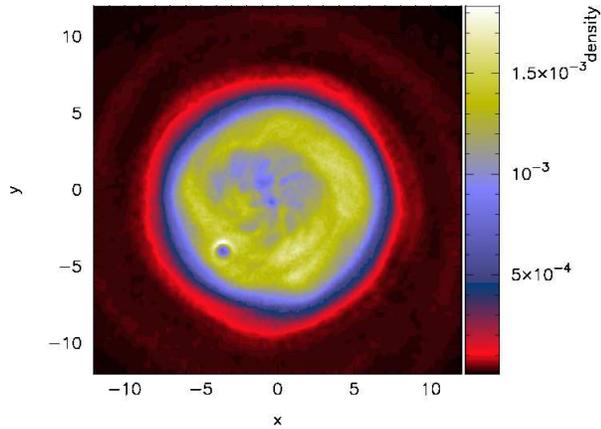, angle=270, scale=0.3}
\caption{Surface density map of the gaseous disk at a particular snapshot of the cosmological run with no winds where the hole is clearly seen at (-4,-4).
The units of density are the same as mentioned in previous surface density plots.}
\label{cosrun}
\end{center}
\end{figure}

A combination of high mass and high velocity is needed to produce holes as shown in the previous section. Due to the different orbital parameters such as angle of impact,
velocity and mass of the halo, the lifetime of the holes is more varied, ranging from a mere $10$ Myr to as much as $70$ Myr. 
Figure \ref{cosrun} shows the surface density map of the gaseous disk after $0.53$ Gyr, a circular hole of about $1$ kpc can be seen in lower 
left corner of the simulation. The structure seen at the center of the galaxy is the result of local instabilities and star formation.

We now quantify how many holes are produced with respect to observations. 
We focus on large holes ($R>1$ kpc), since smaller features can be more easily explained by supernovae explosions, stellar winds or a combination of both.

\begin{figure}
\begin{center}
\epsscale{1.20}
\epsfig{file=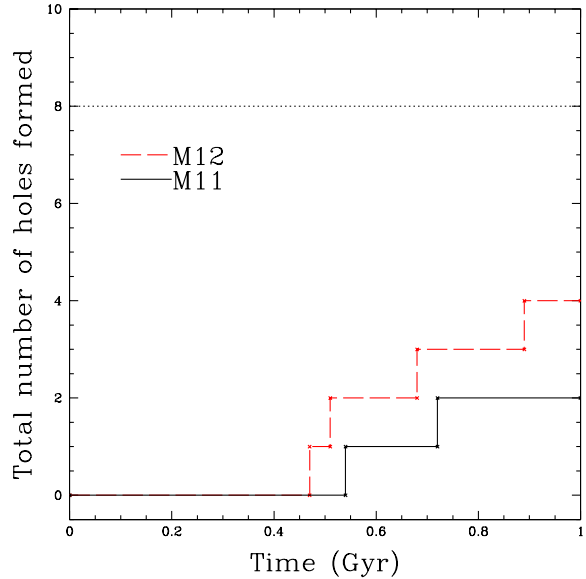, scale=0.4}
\caption{Plot showing the cumulative number of holes formed in a span of $1$ Gyr in the M11 and M12 simulations. The horizontal black dotted line is the number of holes($d>1$ kpc ) 
in one snapshot of IC 2574 \citep{1999AJ....118..273W}. The lower number of holes in the M11 run is, primarily due to its low resolution (compared to M12).} 
\label{events}
\end{center}
\end{figure}

In both our simulation setups (M11 and M12) we observe only $3-4$ events in the $1$ Gyr running time (Figure \ref{events}),
with no more than one hole in a particular snapshot (Figure \ref{cosrun}).
On the other hand \cite{1999AJ....118..273W} observed a total of $8$ holes with radii greater that $1$ kpc in a single ``snapshot'' of IC2574 (dotted line in Figure \ref{events}). 
The number of holes formed in the M11 run is lower primarily due to lack of resolution (when compared to M12). This results in an under estimation of the total number
of low mass satellites.

It emerges clearly that while DM sub-halos are in principle able to create holes in the HI gas, when a cosmologically motivated sub-halo distribution is used
the number of predicted halos do not match the observed one. This leads us to conclude that the sub-halo - disk interaction is not the principal mechanism to 
create the observed HI holes.

We note that the number of holes shown in Figure \ref{events} is just an upper limit. So far we have assumed that DM sub-halos will keep their gas (needed to displace the gas
in the H I disc) all the way down to the galaxy center, completely neglecting the effect of ram pressure.
The ram pressure against the hot gas surrounding the galaxy could partially or totally remove the baryonic content of the sub-halo and make it even more inefficient in producing holes. In order to test this scenario we run an additional simulation, in which we turn on isotropic galactic winds while running the central galaxy in isolation
for $1$ Gyr to stabilize it (see section \ref{sec:sim}).
These winds preferentially expel gas perpendicular to the plane of the disk as it faces the least hindrance in this direction. 
This creates shells of gas surrounding the galactic plane. This gas (slightly colder than the gas inside the sub-halos) is efficient is stripping all the gas from the sub-halos.

When this new initial condition (central galaxy run with winds on) is evolved together with sub-halos $1$ Gyr, it produces virtually no holes.
This experiment shows that it is indeed very difficult for the sub-halos to retain a sufficient amount of gas to perturb the central HI disc. 
This points to a different process being responsible of the production of large holes in the HI distribution.


\section{Conclusions $\&$ Discussion}

Atomic hydrogen (HI) observations of nearby galaxies reveal complex gas distributions. In many systems, the neutral interstellar medium (ISM) contains holes, shells, and/or cavities.
In order to understand the origin of these features, 
we numerically investigate how the impact of dark matter sub-halos orbiting a gas-rich disk galaxy embedded in a massive dark matter halo 
influences the dynamical evolution of the HI gas disk of the galaxy. We mainly focus our attention on the creation of large holes ($R>1$ kpc) in the HI distribution, commonly found in observations of nearby galaxies.

We create a central galaxy resembling the properties of the well-studied dwarf galaxy IC2475 and we bombard its gaseous disk with dark matter dominated satellites. The dark matter, stellar and gas components of both 
the primary galaxy and the satellites are all live (i.e. made of particles) and a full hydro-dynamical code ({\sc gadget2}) including cooling, star formation and feedback is used for these merger simulations.

Our experiment shows that a pure dark matter sub-halo ($M=10^8 M_{\odot}$) is not able to displace the gas in the disk, instead due to gravitational focusing it gives rise to a localized high density region.
This kind of dark impact might be able to induce star formation in the outer part of the extended gas disk, but cannot create large holes.
This result was not unexpected: DM particles interact only through gravity and do not have the necessary contact force to push the gas away from the disk. 

These high density tidal imprints can be characterized and studied in order to obtain the properties of the impacting dark sub-halo as done by \cite{2011arXiv1102.3436C} especially for
relatively massive dark satellites (1:100).

To produce holes we need particles which have contact forces, hence we assume a small amount of gas to be present in the dark matter sub-halo.  
Even a gas fraction as low as $0.8\%$ in a halo of mass $10^8 M_{\odot}$ with a velocity of 150 km/s is able to produce a detectable low density region.
On average holes have a lifetime of about $20-60$ Myr, depending on the halo density, gas mass and impact velocity.

We then use satellite properties directly extracted from high resolution Nbody cosmological simulation (Via Lactea II, Diemand \etal 2008), 
to check how many holes are predicted in the commonly assumed Cold Dark Matter model.

These cosmological motivated runs show that sub-halos with a relatively high gas fraction (5\%), are able to produce a total of about 3-4 large holes ($R>1$ kpc) in an integration time of $1$ Gyr.
This number is significantly lower than the number of observed holes of the same size in IC2475 galaxy. 
If the effect of ram pressure is taken into account, the dark matter satellites tend to loose a significant fraction of their gas content, making them even less efficient 
at perturbing the HI disc.

We conclude that, although DM matter satellites with a modest gas content are in principle able to create holes with a radius of order 1 kpc, disk - sub-halo interaction are 
not the primary channel  through which these holes form in real galaxies. We consider it likely that other astrophysical processes like supernova explosions, stellar winds, high velocity clouds, or a combination of the above factors as the main causes for the observed complex geometry of extended HI disks.

\section*{Acknowledgments}

We thank J\"urg Diemand for making the VLII database publicly available and George Lake for valuable comments. We also thank Adam Leroy for providing the observed HI surface density data of IC2574. 
All numerical simulations used in this work were performed
on the THEO cluster of the Max-Planck-Institut f\"ur Astronomie 
at the Rechenzentrum in Garching. 
Rahul Kannan acknowledges funding by Sonderforschungsbereich SFB 881 
"The Milky Way System" (subproject A1) of the German Research Foundation 
(DFG).

\newpage

\bibliographystyle{apj}
\bibliography{ms}

\end{document}